\pdfoutput=1
\documentclass[a4paper]{article}
\usepackage{INTERSPEECH2022}
\usepackage{amsmath,graphicx,multirow,url,adjustbox}
\usepackage{cite}
\usepackage{color,soul}
\usepackage{hyperref}
\usepackage{mathtools}
\usepackage{tabularx}
\usepackage{booktabs}
\usepackage{stfloats}
\usepackage[pdf]{graphviz}
\usepackage{pgfplots}
\pgfplotsset{width=\linewidth*1.04,compat=1.17}

\usepackage{xpatch}
\makeatletter
\newcommand*{\addFileDependency}[1]{
  \typeout{(#1)}
  \@addtofilelist{#1}
  \IfFileExists{#1}{}{\typeout{No file #1.}}
}
\makeatother
\xpretocmd{\digraph}{\addFileDependency{#2.dot}}{}{}

\newcolumntype{Y}{>{\centering\arraybackslash}X}

\title{CTC Variations Through New WFST Topologies}

\name{\begin{tabular}{c} Aleksandr Laptev$^{1,2}$, Somshubra Majumdar$^1$, Boris Ginsburg$^1$
\thanks{\vspace{-17pt}Preprint. Accepted to Interspeech-22}
\end{tabular}
}
\address{
  $^1$NVIDIA, USA\\
  \vspace{-1pt} 
  $^2$ITMO University, St. Petersburg, Russia}
\email{\{alaptev,smajumdar,bginsburg\}@nvidia.com}

\begin{document}
\maketitle

\begin{abstract}
\vspace{-3pt} 
This paper presents novel Weighted Finite-State Transducer (WFST) topologies to implement Connectionist Temporal Classification (CTC)-like algorithms for automatic speech recognition.
Three new CTC variants are proposed: 
(1) the ``compact-CTC'', in which direct transitions between units are replaced with $\langle \epsilon \rangle$ back-off transitions; 
(2) the ``minimal-CTC'', that only adds $\langle blank \rangle$ self-loops when used in WFST-composition; and 
(3) the ``selfless-CTC'' variants, which disallows self-loop for non-blank units.
Compact-CTC allows for 1.5 times smaller WFST decoding graphs and reduces memory consumption by two times when training CTC models with the LF-MMI objective without hurting the recognition accuracy. Minimal-CTC reduces graph size and memory consumption by two and four times for the cost of a small accuracy drop. Using selfless-CTC can improve the accuracy for wide context window models.
\end{abstract}


\vspace{-3pt} 
\section{Introduction}
\label{sec:intro}
\vspace{-3pt} 

Connectionist Temporal Classification (CTC) \cite{graves_connectionist_2006} is one of the most widely used algorithms to train alignment-free Automatic Speech Recognition (ASR) models. The CTC loss function uses the $\langle blank \rangle$ unit to express the probability of a language unit absence in a particular time-frame. CTC allows repeated language units to indicate that the previous unit continues in the time-frame.

CTC-based ASR models can be combined with an external N-gram Language Model (LM) through a Weighted Finite-State Transducers (WFST) approach. For this end, LM is first compiled into a WFST graph that contains language information. Next, this graph is composed  with a WFST which corresponds to the CTC loss function. The speech decoding is performed through a special type of pruned intersection of an acoustic model's log likelihoods with this composed WFST graph that contains language information and the CTC rules (that  $\langle blank \rangle$ units should be excluded from the recognition results, and repeated units should be collapsed during decoding). Due to the repeated units, the size of the CTC graph increases quadratically with respect to the vocabulary size.  

ASR models with a CTC topology can also be trained with Lattice-Free Maximum Mutual Information (LF-MMI, or just MMI) loss \cite{povey16_interspeech, hadian18_interspeech}. We consider the CTC loss as forward-backward score computation of an intersection between a supervision graph and the acoustic outputs of a neural network. The supervision graph (or numerator graph for MMI) is a composition between a topology graph and a linear graph of the target unit sequence. The MMI loss is expected to deliver better model accuracy than CTC loss \cite{hadian18_interspeech,zhang_benchmarking_lf-mmi} because it directly uses a language model in training: an n-gram LM is used to compute the denominator part of the loss. The WFST representation of the MMI loss also increase quadratically with respect to the  vocabulary size \cite{zheng2021advancing}. If the vocabulary is large (e.g. when the units are word-pieces), this results in a high memory consumption.

Our paper proposes three new CTC-like topologies to (1) decrease WFST decoding graph size for CTC topologies, and (2) reduce memory consumption for CTC models with the MMI loss. We limit our study to CTC-like algorithms that have a unit $\langle blank \rangle$ and allow emitting only one unit per time-frame. Our CTC variants are:
\begin{enumerate}
  \vspace{-4pt} 
  \item The ``compact-CTC'', in which direct transitions between units are replaced with $\langle \epsilon \rangle$ transitions to the initial state. It allows for 1.5 times smaller WFST decoding graphs and reduces GPU memory consumption by two times when training wordpiece-based CTC models with the MMI loss without accuracy reduction compared to the baseline variant.
  \vspace{-4pt} 
  \item The ``minimal-CTC'' endorses the CTC blank-dominant (``peaky'') behavior by removing unit self-loops except for $\langle blank \rangle$, and has no direct transitions between units. It reduces graph size and training memory consumption by two and four times respectively with only ~0.2\% of Word Error Rate (WER) increase.
  \vspace{-4pt} 
  \item The ``selfless-CTC'' modifies variants other than ``minimal-CTC'' by removing self-loops for non-blank units. This topology can improve the accuracy of models with a wide or unrestricted context window.
  \vspace{-4pt} 
\end{enumerate}

This work uses the k2 framework\footnote{\url{https://github.com/k2-fsa/k2}} \cite{povey_k2} for the WFST operations and forward-backward score computation on arbitrary loss variations. The code is released as part of the NeMo toolkit\footnote{\url{https://github.com/NVIDIA/NeMo}}~\cite{kuchaiev2019nemo}.


\begin{figure*}[ht]
\vspace{-27pt} 
\hspace{-12pt} 
\includegraphics[width=1.06\textwidth]{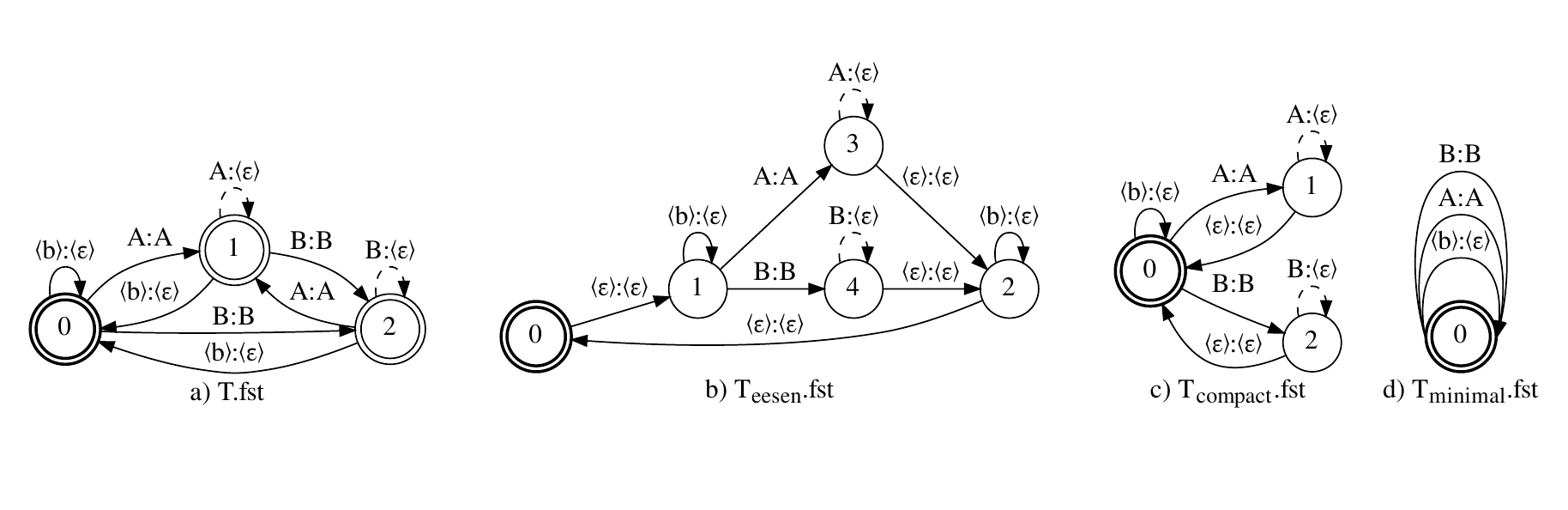}
\vspace{-50pt} 
\caption{Example CTC topologies for a three-unit vocabulary: $\langle blank \rangle$, $A$, and $B$. Variants: a) correct-CTC, b) Eesen-CTC, c) compact-CTC, d) minimal-CTC. $\langle b \rangle$ states for $\langle blank \rangle$. Language unit-to-$\langle \epsilon \rangle$ selfloops are indicated by dashed arrows.}
\label{fig:topologies}
\vspace{-15pt} 
\end{figure*}

\vspace{-3pt} 
\section{Related work}
\label{sec:relwork}
\vspace{-3pt} 

The first two known WFST-based CTC topologies were presented in the Eesen toolkit \cite{miao_eesen} and in \cite{hasim2015}. These variants were suffering from training-inference topology mismatch (decoding topologies did not fully reproduce the training rules). The Eesen topology was discussed in work \cite{xiang_crf}, which proposed a ``correct'' topology and demonstrated a small WER decrease by eliminating the mismatch (see Section~\ref{sec:methods-existing} for more details).

The following papers propose more radical changes to the CTC that are not directly related to the topologies or the WFST apparatus. Zhang et al \cite{zhang20h_interspeech} adapt chenones \cite{le2019senones} (tied context-dependent graphemes) for CTC training, adding an extra state with $\langle blank \rangle$ for each chenone. This modification improves WER over wordpiece-based CTC models, but it limits the model's sub-sampling factor to four. Zhao et al \cite{zhao2019novel} try to decouple the $\langle blank \rangle$ property of filling the space between peaks from modeling silence to get more accurate time segments for units. Works \cite{NEURIPS2018_e44fea3b,LiW20,zeyer2021does} propose CTC modifications to improve language unit time alignment and reduce the peaky behavior. Moritz et al \cite{moritz21_icassp} propose a generalization of CTC that can use ambiguous (non-linear, e.g. graph-based) supervisory information to allow a model to learn the best possible unambiguous target.

\vspace{-3pt} 
\subsection{Existing CTC topologies}
\label{sec:methods-existing}
\vspace{-3pt} 

The default CTC topology in WFST format is presented in Fig.~\hyperref[fig:topologies]{1 a}. We refer to this as T.fst, or the correct-CTC. This is a directed complete graph with self-loops, thus for $N$ units including $\langle blank \rangle$ there are $N$ states and $N^2$ arcs. T.fst can be used for both ASR model training and WFST decoding.

The CTC topology used in Eesen toolkit\cite{miao_eesen} is presented in Fig.~\hyperref[fig:topologies]{1 b}. We refer to this as T\textsubscript{\scalebox{1.2}{eesen}}.fst or the Eesen-CTC. This topology does not allow direct transitions between language units. It has two $\langle blank \rangle$ states which wrap all language units, and $\langle \epsilon \rangle$-linked start state for looping them. This results in $N+2$ states and $3N+1$ arcs for $N$ units. Pure $\langle \epsilon \rangle$ transitions of this graph complicate model training, since the input $\langle \epsilon \rangle$ cannot be aligned with the acoustic outputs. These transitions must be recursively traversed at each time-frame and represented in the resulting lattice to preserve topological order (which is required for some lattice methods, such as forward-backward score computation). Thus T\textsubscript{\scalebox{1.2}{eesen}}.fst should only be used for WFST decoding.

\vspace{-3pt} 
\section{New CTC topologies}
\label{sec:methods-new}
\vspace{-3pt} 

Following the Eesen ideas, we propose a different CTC topology with pure $\langle \epsilon \rangle$ transitions, which is more compact and produces deterministic topologically sorted lattices without significant changes in training pipeline. We call it T\textsubscript{\scalebox{1.2}{compact}}.fst or the ``compact-CTC'' (Fig.~\hyperref[fig:topologies]{1 c}). The topology has only one $\langle blank \rangle$ state, which is the destination for $\langle \epsilon \rangle$ transitions from every language unit. It has $N$ states and $3N-2$ arcs for $N$ units. To be used in training, compact-CTC requires modification of either the acoustic outputs prior to an intersection with the numerator graph or the intersection itself. Here we consider the first option, as it is easier to implement. An additional unit with zero probability is concatenated along the unit dimension of the acoustic output tensor. This augmented tensor is then used to fill the even time-frame slice of the final tensor, which has two times more time-frames and every odd time-frame is filled with probability 1 for each unit. The resulting tensor emulates (with 2X frame redundancy) the $\langle \epsilon \rangle$ transitions back to the $\langle blank \rangle$ state. Although this trick makes the acoustic outputs twice as long, it makes almost no changes to lattice scores. Thus, training with compact-CTC is expected to produce models performing close to the T.fst baseline.

T\textsubscript{\scalebox{1.2}{compact}}.fst can be simplified even more if we remove all language unit-to-$\langle \epsilon \rangle$ self-loops and collapse the $\langle \epsilon \rangle$ transitions. The variant that we call T\textsubscript{\scalebox{1.2}{minimal}}.fst topology or the ``minimal-CTC'' is presented in Fig.~\hyperref[fig:topologies]{1 d}. It has only one state and does only one non-trivial transduction from $\langle blank \rangle$ to $\langle \epsilon \rangle$, resulting in $N$ arcs for $N$ units. T\textsubscript{\scalebox{1.2}{minimal}}.fst can be used for both ASR model training and WFST decoding.

The idea of removing language unit-to-$\langle \epsilon \rangle$ self-loops can also be applied to any of the aforementioned topologies except T\textsubscript{\scalebox{1.2}{minimal}}.fst, where there are no such self-loops by design. We call this variation ``selfless-CTC''. Variants of selfless-CTC can be obtained from Fig.~\ref{fig:topologies} by removing dashed transitions. Note that while selfless T\textsubscript{\scalebox{1.2}{compact}}.fst and T\textsubscript{\scalebox{1.2}{minimal}}.fst are used differently in training (T\textsubscript{\scalebox{1.2}{minimal}}.fst does not require acoustic output manipulations performed for T\textsubscript{\scalebox{1.2}{compact}}.fst), removing self-loops from T\textsubscript{\scalebox{1.2}{compact}}.fst makes it almost identical to T\textsubscript{\scalebox{1.2}{minimal}}.fst from the the Kaldi \cite{Povey11thekaldi} WFST decoding perspective ($\langle \epsilon \rangle$ transitions are processed in a breadth-first manner in a single decoding iteration; this leads to almost identical unit pools when using graphs differing only in $\langle \epsilon \rangle$ transitions). Decoding with regular T\textsubscript{\scalebox{1.2}{compact}}.fst is also expected to be identical to using T\textsubscript{\scalebox{1.2}{eesen}}.fst, thus the topology that results in a smaller decoding graph should be preferred for use.

\vspace{-3pt} 
\section{Experiments}

\vspace{-3pt} 
\subsection{Experimental setup}
\vspace{-3pt} 

We used the NeMo toolkit for ASR training, k2 framework to compute loss functions on arbitrary topologies, and Riva ASR Library\footnote{\url{https://github.com/nvidia-riva/riva-asrlib-decoder}} for GPU WFST decoding \cite{cuda_decoder}. Alternatively, k2 can also be used for the decoding, as well as the modified kaldi decoder\footnote{\url{https://github.com/thu-spmi/CAT/blob/master/src/kaldi-patch/latgen-faster.cc}} used in \cite{xiang_crf}.

We considered three ASR models: Citrinet-384, Citrinet-1024 \cite{majumdar2021citrinet}, and Conformer-L \cite{Gulati2020}.  Hyperparameters are taken as in the corresponding NeMo configuration files.  Experiments were performed with the 3-way speed-perturbed LibriSpeech \cite{panayotov2015librispeech}, and the 4-gram word-level LM provided with the dataset. The following settings are used by default unless otherwise noted:
\begin{itemize}
    \vspace{-4pt} 
    \item The default ASR model is Citrinet-384 with context scaling factor $\gamma=0.25$, trained for 400 epochs on 32 V100 GPUs, and with 256 Byte-Pair Encoding (BPE) wordpiece vocabulary. CTC models were trained with batch size 32 per GPU.
    \vspace{-4pt} 
    \item MMI models were trained with bi-gram unit-level LMs without backoff probabilities. Decoding graphs for MMI models include the unit-level LM used in training, which is composed with the topology graph.  MMI models were first trained with batch 16 per GPU and gradient accumulation 2 for 4 epochs, then with batch 32 for the rest of the training (see Section~\ref{sec:supplementary-study}).
    \vspace{-4pt} 
    \item WER results were obtained with WFST decoding on a graph constructed with the same topology used to train the particular model.
    \vspace{-4pt} 
\end{itemize}

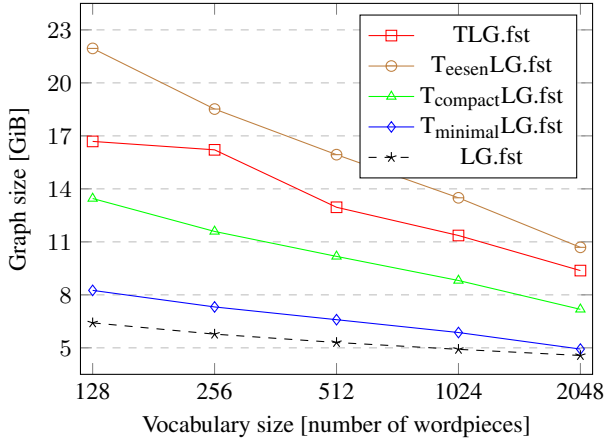
\begin{figure}[t]
\hspace{-3pt}
\begin{tikzpicture}
\pgfplotsset{%
    x tick label style={/pgf/number format/1000 sep=\,},
    log base 10 number format code/.code={%
        $\pgfmathparse{10^(#1)}\pgfmathprintnumber{\pgfmathresult}$%
    }%
}  
\begin{axis}[
    xmode=log,
    log ticks with fixed point,
    xlabel={Vocabulary size [number of wordpieces]},
    ylabel={Graph size [GiB]},
    height=6.5cm,
    xmin=128, xmax=2048,
    ymin=4, ymax=24,
    xtick={128,256,512,1024,2048},
    xticklabels={128,256,512,1024,2048}, 
    ytick={5,8,11,14,17,20,23},
    legend pos=north east,
    ymajorgrids=true,
    grid style=dashed,
	enlargelimits=0.025,
]

\addplot[
    color=red,
    mark=square,
    ]
    coordinates {
    (128,16.686580021)(256,16.210114081)(512,12.958632869)(1024,11.360642777)(2048,9.374569253)};
    \addlegendentry{TLG.fst}

\addplot[
    color=brown,
    mark=halfcircle,
    ]
    coordinates {
    (128,21.951218701)(256,18.518031621)(512,15.936703145)(1024,13.494135421)(2048,10.690085437)};
    \addlegendentry{T\textsubscript{\scalebox{1.2}{eesen}}LG.fst}

\addplot[
    color=green,
    mark=triangle,
    ]
    coordinates {
    (128,13.454113621)(256,11.590339129)(512,10.177699053)(1024,8.813753657)(2048,7.180788233)};
    \addlegendentry{T\textsubscript{\scalebox{1.2}{compact}}LG.fst}

\addplot[
    color=blue,
    mark=diamond,
    ]
    coordinates {
    (128,8.253983465)(256,7.316065853)(512,6.594173965)(1024,5.867700781)(2048,4.931680801)};
    \addlegendentry{T\textsubscript{\scalebox{1.2}{minimal}}LG.fst}

\addplot[
    color=black,
    mark=star,
    dashed]
    coordinates {
    (128,6.414693433)(256,5.775879145)(512,5.303925905)(1024,4.915517181)(2048,4.568019881)};
    \addlegendentry{LG.fst}

\end{axis}
\end{tikzpicture}
\vspace{-20pt} 
\caption{Decoding graph sizes with the considered topologies.}
\label{fig:graph-plot}
\vspace{-15pt} 
\end{figure}


\begin{table}[b!]
\centering
\vspace{-15pt} 
\caption{CTC-trained vs MMI-trained variations, WER [\%]}
\label{tab:Citrinet-384-ctc-vs-mmi}
\vspace{-10pt} 
\begin{tabularx}{\linewidth}{c|c|YYYY} 
 \toprule
 \multirow{2}{*}{\textbf{Loss}}       &
 \multirow{2}{*}{\textbf{CTC variants}}   &
 \multicolumn{2}{c}{\textbf{Dev}}     &
 \multicolumn{2}{c}{\textbf{Test}}    \\
  &&\textbf{clean}& \textbf{other}& \textbf{clean}& \textbf{other}  \\
\midrule
 \multirow{5}{*}{CTC} & correct & 2.57 & 6.82 & 2.91 & 6.98 \\
  & correct selfless            & 3.52 & 9.10 & 3.72 & 9.61 \\
  & compact                     & na & na & na & na \\
  & compact selfless            & 3.69 & 9.65 & 3.95 & 9.89 \\
  & minimal                     & 3.74 & 9.52 & 3.99 & 10.07 \\
\midrule
 \multirow{5}{*}{MMI} & correct & 2.65 & 6.64 & 2.91 & 6.82 \\
  & correct selfless            & 2.66 & 6.80 & 2.96 & 6.94 \\
  & compact                     & 2.66 & 6.62 & 2.98 & 6.83 \\
  & compact selfless            & 2.80 & 7.14 & 3.14 & 7.45 \\
  & minimal                     & 2.85 & 7.14 & 3.01 & 7.36 \\
 \bottomrule
\end{tabularx}
\vspace{-5pt} 
\end{table}

\vspace{-3pt} 
\subsection{Preliminary study}
\label{sec:preliminary-study}
\vspace{-3pt} 

A decoding graph consists of a topology graph, a lexicon graph ($L.fst$) that maps vocabulary unit sequences to words, and a word-level LM graph ($G.fst$, or grammar) composed together. Fig.~\ref{fig:graph-plot} shows a comparison of different decoding graph sizes against different vocabulary sizes for the correct-CTC, Eesen-CTC, compact-CTC, and minimal-CTC. Selfless-CTC variants are not shown for plot compactness (their graph sizes are almost the same as of their base variants with a constant decrease caused by fewer self-loops). $G.fst$ is the same for every setup. TLG.fst sets the baseline, and the language graph LG.fst serves as the lower bound of the size a decoding graph can have (composition with a topology graph can only enlarge the final graph). T\textsubscript{\scalebox{1.2}{eesen}}LG.fst has larger sizes than the baseline while using T\textsubscript{\scalebox{1.2}{compact}}LG.fst reduces memory consumption by a quarter. T\textsubscript{\scalebox{1.2}{minimal}}.fst gives the least topology overhead because it introduces only the $\langle blank \rangle$ self-loops when composed with LG.fst. Based on this data and the above discussion on $\langle \epsilon \rangle$ transitions, we exclude T\textsubscript{\scalebox{1.2}{eesen}}LG.fst and selfless-T\textsubscript{\scalebox{1.2}{compact}}LG.fst from consideration in this work. At the same time, selfless-TLG.fst is of interest because it cannot be fully emulated using any other topology, and its size is smaller than that of the baseline TLG.fst.

Note that for a word-level LM, the graph size decreases as the vocabulary size increases (for a decoding graph TG.fst with wordpiece-based G.fst we would observe a directly proportional dependence of the sizes). We believe there are two reasons for this. First, unit sequences in L.fst become shorter with the vocabulary size increase, which results in smaller L.fst size. Secondly, at the composition of a topology and LG.fst, at almost every state most of the arcs of T.fst are not used (because L.fst is a union of linear graphs representing unit sequences, and even after graph minimization, the average number of arcs per state is less than 5 for most graphs). All this takes precedence over the topology size increase. Thus, if WER is close to optimal for some of the unit vocabularies, the largest one should be preferred.

The initial experiments were to evaluate the CTC's ability to operate effectively with the alternative topologies that can be used in training. They are: T.fst and T\textsubscript{\scalebox{1.2}{compact}}.fst with or without self-loops, and T\textsubscript{\scalebox{1.2}{minimal}}.fst. The first part of Table~\ref{tab:Citrinet-384-ctc-vs-mmi} contains WER results of purely CTC-trained models with different topologies. Whilst with the CTC loss all the proposed variations demonstrated a significant accuracy drop against correct-CTC as a baseline (more over, compact-CTC did not converge), with the MMI loss all the variations demonstrate close results (the second part of Table~\ref{tab:Citrinet-384-ctc-vs-mmi}). In the rest of the paper, we explore training with different topologies with the MMI loss only, keeping  CTC-trained correct-CTC models as references.

\begin{table}[t]
\centering
\vspace{-5pt} 
\caption{Citrinet-1024, variants with self-loops, WER [\%]}
\label{tab:Citrinet-1024}

\vspace{-10pt} 
\begin{tabularx}{\linewidth}{c|c|YYYY} 
 \toprule
 \multirow{2}{*}{\textbf{Loss}}      &
 \multirow{2}{*}{\textbf{\shortstack{Decoding \\ approach}}}    &
 \multicolumn{2}{c}{\textbf{Dev}}    &
 \multicolumn{2}{c}{\textbf{Test}}   \\
  &&\textbf{clean}& \textbf{other}& \textbf{clean}& \textbf{other}  \\
\midrule
 \multirow{3}{*}{CTC} & greedy    & 2.38 & 6.00 & 2.57 & 5.87 \\
  & TLG.fst                           & \textbf{2.12} & 4.83 & 2.40 & 5.12 \\
  & T\textsubscript{\scalebox{1.2}{compact}}LG.fst & \textbf{2.12} & 4.83 & 2.39 & 5.13 \\
\midrule
 \multirow{3}{*}{MMI} & greedy    & 2.16 & 5.55 & 2.36 & 5.35 \\
  & TLG.fst                           & 2.14 & 4.78 & 2.40 & \textbf{4.89} \\
  & T\textsubscript{\scalebox{1.2}{compact}}LG.fst & 2.14 & 4.78 & 2.39 & \textbf{4.89} \\
\midrule
 \multirow{3}{*}{\shortstack{MMI \\ compact}} 
  & greedy                        & 2.27 & 5.34 & \textbf{2.31} & 5.32 \\
  & TLG.fst                           & 2.23 & 4.55 & 2.43 & 4.93 \\
  & T\textsubscript{\scalebox{1.2}{compact}}LG.fst & 2.22 & \textbf{4.50} & 2.41 & 4.93 \\
 \bottomrule
\end{tabularx}
\vspace{-15pt} 
\end{table}

\begin{table}[b!]
\centering
\vspace{-15pt} 
\caption{Citrinet-1024, selfless-CTC, WER [\%]}
\label{tab:Citrinet-1024-no-self-loops}
\vspace{-10pt} 
\begin{tabularx}{\linewidth}{c|c|YYYY} 
 \toprule
 \multirow{2}{*}{\textbf{Loss}}      &
 \multirow{2}{*}{\textbf{\shortstack{Decoding \\ approach}}}    &
 \multicolumn{2}{c}{\textbf{Dev}}    &
 \multicolumn{2}{c}{\textbf{Test}}   \\
  &&\textbf{clean}& \textbf{other}& \textbf{clean}& \textbf{other}  \\
\midrule
 \multirow{3}{*}{MMI} & greedy     & 2.22 & 5.42 & \textbf{2.43} & 5.32 \\
  & TLG.fst                            & \textbf{2.11} & \textbf{4.64} & \textbf{2.43} & \textbf{4.91} \\
  & T\textsubscript{\scalebox{1.2}{minimal}}LG.fst & 2.23 & 4.76 & 2.53 & 5.07 \\
\midrule
 \multirow{2}{*}{\shortstack{MMI \\ compact}} 
  & greedy                     & 2.39 & 6.22 & 2.64 & 6.13 \\
  & T\textsubscript{\scalebox{1.2}{minimal}}LG.fst & 2.22 & 5.16 & 2.56 & 5.33 \\
\midrule
 \multirow{2}{*}{\shortstack{MMI \\ minimal}}
  & greedy                     & 2.41 & 6.53 & 2.71 & 6.25 \\
  & T\textsubscript{\scalebox{1.2}{minimal}}LG.fst & 2.27 & 5.11 & 2.64 & 5.36 \\
 \bottomrule
\end{tabularx}
\vspace{-5pt} 
\end{table}

\vspace{-3pt} 
\subsection{The MMI models with different CTC variants}
\label{sec:main-study}
\vspace{-3pt} 

We considered two ASR models: Citrinet-1024 and Conformer-L, trained for 1k epochs. WER is measured with five decoding approaches: greedy (usual argmax for CTC models and ``argmax of pairs'' -- intersection with the bi-gram training LM for MMI models), and WFST decoding with TLG.fst, selfless-TLG.fst, T\textsubscript{\scalebox{1.2}{compact}}LG.fst, and T\textsubscript{\scalebox{1.2}{minimal}}LG.fst. Only applicable decoding approaches are demonstrated for each model. In Tables~\ref{tab:Citrinet-1024}--\ref{tab:Conformer-L-no-self-loops}, values in Loss column state as follows: ``CTC'' for a CTC-trained model with T.fst topology, ``MMI'' for an MMI-trained model with T.fst, ``MMI compact'' for T\textsubscript{\scalebox{1.2}{compact}}.fst, and ``MMI minimal'' for T\textsubscript{\scalebox{1.2}{minimal}}.fst.

Tables~\ref{tab:Citrinet-1024} and~\ref{tab:Citrinet-1024-no-self-loops} demonstrate results for Citrinet-1024 and variants with- and without self-loops, respectively. The MMI models, while being generally better, enjoyed lesser accuracy improvement than the CTC model from decoding with WFST graph versus greedy mode. Decoding with T\textsubscript{\scalebox{1.2}{compact}}LG.fst is preferred over TLG.fst because it allows for 1.5 times graph size reduction for the cost of speed reduction by 5-10\% without WER increase. Compact-CTC and selfless-CTC models performed close to the correct-CTC. Moreover, the WER increase due to topology mismatch when using T\textsubscript{\scalebox{1.2}{minimal}}LG.fst with the correct-selfless-CTC did not exceed 0.2\%. This might be a reasonable tradeoff because the size of T\textsubscript{\scalebox{1.2}{minimal}}LG.fst and the decoding time are nearly 2 times smaller than that of TLG.fst. Compact-selfless-CTC and minimal-CTC performed worse than the CTC baseline.

\begin{table}[t!]
\centering
\vspace{-5pt} 
\caption{Conformer-L, variants with self-loops, WER [\%]}
\label{tab:Conformer-L}
\vspace{-10pt} 
\begin{tabularx}{\linewidth}{c|c|YYYY} 
 \toprule
 \multirow{2}{*}{\textbf{Loss}}      &
 \multirow{2}{*}{\textbf{\shortstack{Decoding \\ approach}}}    &
 \multicolumn{2}{c}{\textbf{Dev}}    &
 \multicolumn{2}{c}{\textbf{Test}}   \\
  &&\textbf{clean}& \textbf{other}& \textbf{clean}& \textbf{other}  \\
\midrule
 \multirow{3}{*}{CTC} & greedy    & 2.52 & 6.50 & 2.94 & 6.65 \\
  & TLG.fst                           & 2.27 & 5.11 & 2.68 & \textbf{5.35} \\
  & T\textsubscript{\scalebox{1.2}{compact}}LG.fst & 2.28 & 5.10 & 2.67 & \textbf{5.35} \\
\midrule
 \multirow{3}{*}{MMI} & greedy    & 2.58 & 6.35 & 2.69 & 6.48 \\
  & TLG.fst                           & 2.32 & 5.20 & 2.61 & 5.58 \\
  & T\textsubscript{\scalebox{1.2}{compact}}LG.fst & 2.32 & 5.20 & 2.61 & 5.56 \\
\midrule
 \multirow{3}{*}{\shortstack{MMI \\ compact}} 
  & greedy                        & 2.42 & 6.39 & 2.62 & 6.51 \\
  & TLG.fst                           & \textbf{2.23} & 5.10 & \textbf{2.55} & 5.58 \\
  & T\textsubscript{\scalebox{1.2}{compact}}LG.fst & 2.24 & \textbf{5.09} & \textbf{2.55} & 5.56 \\
 \bottomrule
\end{tabularx}
\vspace{-15pt} 
\end{table}

\begin{table}[b]
\centering
\vspace{-15pt} 
\caption{Conformer-L, selfless-CTC, WER [\%]}
\label{tab:Conformer-L-no-self-loops}
\vspace{-10pt} 
\begin{tabularx}{\linewidth}{c|c|YYYY} 
 \toprule
 \multirow{2}{*}{\textbf{Loss}}      &
 \multirow{2}{*}{\textbf{\shortstack{Decoding \\ approach}}}    &
 \multicolumn{2}{c}{\textbf{Dev}}    &
 \multicolumn{2}{c}{\textbf{Test}}   \\
  &&\textbf{clean}& \textbf{other}& \textbf{clean}& \textbf{other}  \\
\midrule
 \multirow{3}{*}{MMI} & greedy     & 2.56 & 6.49 & 2.80 & 6.48 \\
  & TLG.fst                            & \textbf{2.23} & \textbf{4.99} & \textbf{2.54} & \textbf{5.33} \\
  & T\textsubscript{\scalebox{1.2}{minimal}}LG.fst & 2.34 & 5.11 & 2.68 & 5.50 \\
\midrule
 \multirow{2}{*}{\shortstack{MMI \\ compact}} 
  & greedy                     & 3.09 & 8.24 & 3.39 & 7.99 \\
  & T\textsubscript{\scalebox{1.2}{minimal}}LG.fst & 2.50 & 5.91 & 2.82 & 6.01 \\
\midrule
 \multirow{2}{*}{\shortstack{MMI \\ minimal}}
  & greedy                     & 3.24 & 8.64 & 3.42 & 8.50 \\
  & T\textsubscript{\scalebox{1.2}{minimal}}LG.fst & 2.55 & 6.17 & 2.80 & 6.28 \\
 \bottomrule
\end{tabularx}
\vspace{-5pt} 
\end{table}

In contrast to Citrinet-1024, Conformer-L (trained with 128 BPE wordpieces, and with batch 16 on 128 V100 GPUs) did not benefit much from using MMI (Tables~\ref{tab:Conformer-L} and~\ref{tab:Conformer-L-no-self-loops}). Only compact-selfless-CTC slightly outperformed the CTC baseline. Other decoding-related observations are consistent with those obtained earlier for Citrinet-1024.

\vspace{-3pt} 
\subsection{The effect of context and vocabulary sizes on models with different topologies.}
\label{sec:supplementary-study}
\vspace{-3pt} 

Since the best models for our Conformer (unlimited context) and Citrinet ($\gamma=0.25$, 11 frames of context) belong to different self-loop variants, we compared two Citrinets with different context sizes ($\gamma=0.25$ and $\gamma=1.0$). Compact-selfless-CTC and minimal-CTC models are excluded from this consideration, as they are less effective than the other variants. According to Table~\ref{tab:Citrinet-384-kernel-sizes}, variants with self-loops performed better than without for the short context. However, for the longer context these variants show some WER increase, while selfless-CTC improved the accuracy and delivered the best result overall. This finding is consistent with the Conformer results in Section~\ref{sec:main-study}. Removing self-loops from the CTC topology can be beneficial for ASR models with long contexts.

LF-MMI models are usually trained on chunks of 1 to 1.5 seconds in Kaldi and PyChain \cite{shao20_interspeech}, while we trained our models on whole utterances (since there are no reference alignments for CTC-MMI models), which are 10 to 15 times longer. This, and the uniform distribution of the log-probabilities at the start of the training, leads to a sharp increase in the denominator lattice size and can cause Out Of Memory (OOM) errors. We observed that after the first 2-4 epochs of training (which can be treated as a warmup) the log-probability distribution becomes less uniform and most of the arcs can be pruned out during the lattice generation without affecting model convergence. To train MMI models with BPE units, we had to reduce batch size (and increase gradient accumulation accordingly) during the warmup. After the warmup, the MMI loss calculation requires no more than a few hundred MB of memory and slows down the training by only 2-5\%. Table~\ref{tab:Citrinet-384-vocabulary-sizes} demonstrates the maximum initial batch that does not cause OOM for models with different topologies relative to the BPE vocabulary size. The most memory efficient topology is T\textsubscript{\scalebox{1.2}{minimal}}.fst, which allows to increase batch size up to four times, and T\textsubscript{\scalebox{1.2}{compact}}.fst provides a steady memory consumption reduction (up to two times) against T.fst.

\begin{table}[t!]
\centering
\vspace{-5pt} 
\caption{Variations against different context sizes, WER [\%]}
\label{tab:Citrinet-384-kernel-sizes}
\vspace{-10pt} 
\begin{tabularx}{\linewidth}{c|c|YYYY} 
 \toprule
 \multirow{2}{*}{\textbf{Gamma}}       &
 \multirow{2}{*}{\textbf{CTC variants}}   &
 \multicolumn{2}{c}{\textbf{Dev}}     &
 \multicolumn{2}{c}{\textbf{Test}}    \\
  &&\textbf{clean}& \textbf{other}& \textbf{clean}& \textbf{other}  \\
\midrule
 \multirow{3}{*}{0.25} & correct & 2.65 & 6.64 & 2.91 & 6.82 \\
  & correct selfless             & 2.66 & 6.80 & 2.96 & 6.94 \\
  & compact                      & 2.66 & 6.62 & 2.98 & 6.83 \\
\midrule
 \multirow{3}{*}{1.0} & correct & \textbf{2.58} & 6.61 & 2.99 & 6.86 \\
  & correct selfless            & 2.60 & \textbf{6.43} & \textbf{2.86} & \textbf{6.64} \\
  & compact                     & 2.68 & 6.81 & 3.06 & 7.01 \\
 \bottomrule
\end{tabularx}
\vspace{-15pt} 
\end{table}

\begin{table}[!ht]
\centering
\vspace{-5pt} 
\caption{Maximum initial batch.}
\label{tab:Citrinet-384-vocabulary-sizes}
\vspace{-10pt} 
\setlength\tabcolsep{4.5pt} 
\begin{tabularx}{\linewidth}{Y|c|c|c|c|c} 
 \toprule
 \multirow{3}{*}{\textbf{\shortstack{Voc.\\size}}} &
 \multicolumn{5}{c}{\textbf{CTC variants}}  \\ \cline{2-6} \rule{0pt}{10pt}
  & \multirow{2}{*}{\textbf{correct}} & \multirow{2}{*}{\textbf{\shortstack{correct\\selfless}}} & \multirow{2}{*}{\textbf{compact}} & \multirow{2}{*}{\textbf{\shortstack{compact\\selfless}}} & \multirow{2}{*}{\textbf{minimal}} \\
 &&&&& \\
\midrule
 256 & 16 & 16 & 32 & 32 & 32 \\
 512 & 8 & 4 & 16 & 16 & 16 \\
 1024 & 4 & 4 & 8 & 8 & 8 \\
 2048 & 1 & 1 & 2 & 2 & 4 \\
 \bottomrule
\end{tabularx}
\vspace{-15pt} 
\end{table}

\section{Conclusion}
\vspace{-3pt} 

This paper presented a thorough study of novel WFST graph representations (topologies) for CTC-like algorithms for training and inference of ASR models. We analyzed the existing CTC topologies and proposed three new CTC-like variants to reduce graph sizes for WFST-based decoding and memory consumption for training CTC models with the MMI loss.
The ``Compact-CTC'' topology and corresponding traversal algorithm replaces direct transitions between units with $\langle \epsilon \rangle$ back-off transitions. The WFST graph of this variant is fully compatible with the classic (correct) CTC and provides the same decoding accuracy with almost the same speed while occupying 1.5 times less memory.
The ``Minimal-CTC'' topology only adds $\langle blank \rangle$-to-$\langle \epsilon \rangle$ self-loops when used in WFST-composition. The decoding graph for this variant has the minimal topology-induced overhead, and its size is half that of the base graph. The decoding speed is also two times faster for the cost of ~0.2\% WER increase.
The ``Selfless-CTC'' variation disallows language unit-to-$\langle \epsilon \rangle$ self-loops for any CTC variant except minimal-CTC. Any selfless-CTC variant is compatible with the minimal-CTC graph at decoding. Also, selfless-CTC for the correct topology can reduce WER for wide context window models.
Compact-CTC and minimal-CTC require two and four times less GPU memory, respectively, compared to the correct one, allowing wordpiece-based models with a large vocabulary to be trained with the LF-MMI objective.

\vspace{-3pt} 
\section{Acknowledgements}
\vspace{-3pt} 

We would like to thank V. Noroozi, V. Lavrukhin, D. Galvez, E. Rastorgueva, and the k2-FSA team for advice and support.
\bibliography{refs}
\end{document}